\documentclass[twocolumn,english,prb,showpacs]{revtex4-1}
\usepackage{hyperref}
\usepackage{natbib}
\usepackage[latin9]{inputenc}
\usepackage{amstext,bm}
\usepackage{lipsum}
\usepackage{graphicx}
\usepackage{xcolor}
\usepackage{amsmath}
\usepackage{babel}
\usepackage[titletoc]{appendix}
\usepackage{float}
\usepackage{longtable}

\begin{document}

\title{{Exotic Node Line in ternary MgSrSi-type crystals}}

\author{Jinling Lian$^1$} \author{Lixian Yu$^1$}\author{Qi-Feng Liang$^{1}$}\email{qfliang@usx.edu.cn} \author{Jian Zhou$^{2}$} \author{Rui Yu$^{3}$}\email{ruiyu@whu.edu.cn} \author{Hongming Weng$^{5}$}
\address{$^{1}$ Department of Physics, Shaoxing University, Shaoxing 312000, P. R. China}
\address{$^{2}$ National Laboratory of Solid State Microstructures and Department of Materials Science and Engineering, Nanjing University, Nanjing 210093, China}
\address{$^{3}$ School of Physics and Technology, Wuhan University, Wuhan 430072, P. R. China}
\address{$^{4}$ Beijing National Laboratory for Condensed Matter Physics, and Institute of Physics, Chinese Academy of Sciences,Beijing 100190,  P. R. China}

\begin{abstract}
Node line (NL) band-touchings protected by mirror symmetry (named as $m$-NL), the product of inversion and time reversal symmetry $\mathcal{S=PT}$ ($s$-NL), or nonsymmorphic symmetry are nontrivial topological objects of topological semimetals (SMs) in the Brillouin Zone (BZ). In this work, we screened the family of MgSrSi-type crystals using first principles calculations, and discovered that more than 70 members are NLSMs. A new type of multi-loop structure was found in AsRhTi that a $s$-NL touches robustly with a $m$-NL at some ``nexus point", and in the meanwhile a second $m$-NL crosses with the $s$-NL to form a Hopf-link. Unlike the previously proposed Hopf-link formed by two $s$-NLs or two $m$-NLs, a Hopf-link formed by a $s$-NL and a $m$-NL requires a minimal three-band model to characterize its essential electronic structure. The associated topological surface states on different surfaces of AsRhTi crystal were also obtained. Even more complicated and exotic multi-loop structure of NLs were predicted in AsFeNa and PNiNb. Our work may shed light on search for exotic multi-loop NLSMs in real materials.
\end{abstract}
\maketitle

\section{Introduction}
The band crossings of the conduction and valence bands in a topological semimetal are interesting topological objects of Brillouin Zone (BZ) which exhibit unique electronic structures and electrical properties, such as giant magnetoresistance, parity anomaly and ``drum-head" states at systems' surfaces.\cite{TSM_WengJPCM2016,TSM_FangCPB2016} Depending on the dimensionality of band crossings, topological semimetals are classified into three categories, the Weyl semimetals (WSMs)\cite{WSM_Wan,WSM_Burkov,WSM_Xu,WSM_Weng,WSM_Lv,WSM_XSY}or Dirac semimetals (DSMs),\cite{DSM_Murakami,DSMNa3Bi,DSMCd3As2} node-line semimetals (NLSMs), \cite{NLSM_BurkovPRB2011,NLSM_WengPRB2015,NLSM_KimPRL2015,NLSM_YuPRL2015,NLSM_XieAPLM2015,NLSM_BianPRB2016,NLSM_BianNC2016,Heikkila_arxiv2015,
NLSM_FangCPRB2015,NLSM_PhillipsPRB2014,NLSM_WeberPRB2017,NLSM_BePRL2016,NLSM_ZrSiSPRB2016,NLSM_ZrSiSePRL2016,NLSM_NodeChainNature2016,
NLSM_YuHfCPRL2017,NLSM_CaP3PRB2017,NLSM_CaAs3PRL2017,NLSM_WangHopfPRB2017,NLSM_ChenHopfPRB2017,NLSM_CoMnGaPRL2017,NLSM_AnHopfPRB2018,NLSM_NNetPRL2018,
NLSM_WTLHfPtGe2018} and node-surface semimetals (NSSMs).\cite{NSSM_LiangPRB2016,NSSM_arxiv2018} Unlike the DSMs and WSMs where the band crossings take place at discrete points in the BZ, the band crossings of a NLSM form closed loops. When circles around these loops, an electron picks up an nontrivial Berry phase $\pi$ in its wave function, whose effect can be detected by transporting measurements. Though have being extensively proposed in graphene networks\cite{NLSM_WengPRB2015}, anti-perovskites,\cite{NLSM_KimPRL2015,NLSM_YuPRL2015} SrIrO$_3$,\cite{NLSM_FangCPRB2015} TlTaS$_2$,\cite{NLSM_BianPRB2016} BaTaS,\cite{NSSM_LiangPRB2016} HfC,\cite{NLSM_YuHfCPRL2017} CaP$_3$/CaAs$_3$\cite{NLSM_CaP3PRB2017,NLSM_CaAs3PRL2017} and Co$_2$MnGa,\cite{NLSM_CoMnGaPRL2017} etc, the direct evidence of existence of NLSMs states in real materials are rare. \cite{NLSM_BianNC2016,NLSM_BePRL2016,NLSM_ZrSiSPRB2016,NLSM_ZrSiSePRL2016} Finding new materials with clean and robust NL band crossing around the fermi level is still a demanding task in the field of condensed matter physics.

\begin{figure*}[htbp]
\centering{}
\includegraphics[width=0.95\textwidth]{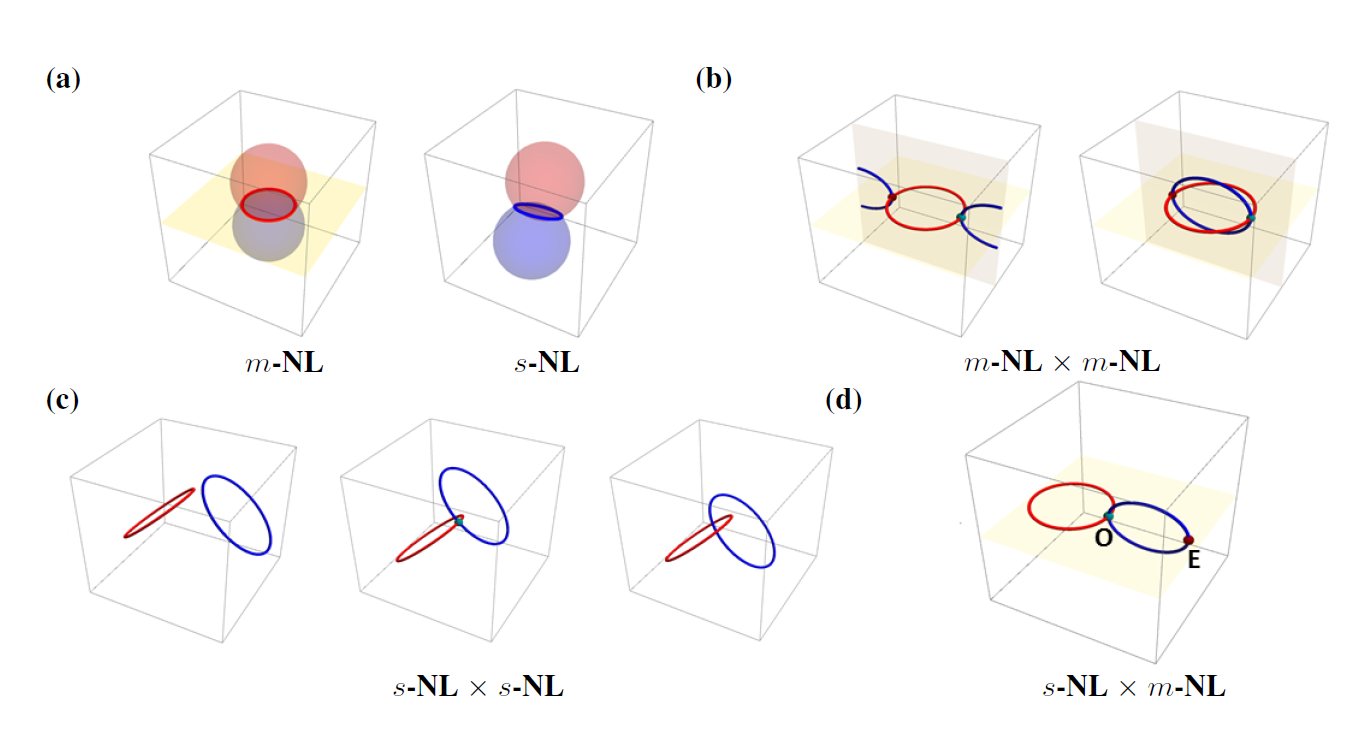}
\caption{(Color online)
Possible NL structures in BZ. (a) $m$-NL protected by mirror symmetry (left) and a $s$-NL protected by symmetry $\mathcal{S=PT}$ (right). (b) A pair of vertical and touched $m$-NLs associated with two vertical mirror planes. The small spheres are located at the touching points of two NLs is also dubbed as "nexus points". (c) A pair of separated, touched and crossed $s$-NLs depending on their distance in the BZ. (d) A touched $s$-NL and $m$-NL. Except the nexus point O, the $s$-NL also penetrates the invariant plane of the $m$-NL through another isolated point marked as E.
\label{fig:NL}}
\end{figure*}

Three types of NLs have been discovered based on their protecting symmetry.\cite{TSM_FangCPB2016} The first type of NLs are those protected by mirror symmetries, which are named as $m$-NLs in this work and shown schematically in Fig. (\ref{fig:NL}a). Due to the mirror symmetry, a $m$-NL is pinned to the invariant plane of the mirror symmetry. The second type of NLs  are consisted of NLs protected by the combination of time-reversal symmetry $\mathcal{T}$ and inversion symmetry $\mathcal{P}$, i.e. $\mathcal{S=PT}$. This type of NLs, named as $s$-NLs here, can present at any region of the BZ as shown at the right of Fig. (\ref{fig:NL}a). The last type of NLs are protected by nonsymmorphic symmetries and usually appear at the boundary of BZ.\cite{NLSM_FangCPRB2015,NSSM_LiangPRB2016}

Recently, there rises a new trend of investigating NLSMs with multiple NL loops.\cite{NLSM_NodeChainNature2016,NLSM_YuHfCPRL2017,NLSM_WangHopfPRB2017,NLSM_ChenHopfPRB2017,NLSM_CoMnGaPRL2017,NLSM_AnHopfPRB2018,NLSM_NNetPRL2018} In those NLSMs, NL loops may intersect with each other and entangle into a variety of structures, such as node-net,\cite{NLSM_NNetPRL2018} node-chain\cite{NLSM_NodeChainNature2016} and Hopf-link,\cite{NLSM_WangHopfPRB2017,NLSM_ChenHopfPRB2017,NLSM_CoMnGaPRL2017,NLSM_AnHopfPRB2018} etc.
For example, two $m$-NLs will be stuck together at some points dubbed as ``nexus points" on the cross-line of two invariant planes of mirror symmetries (see in Fig.(\ref{fig:NL}b)).\cite{NLSM_NexusNJP2015} In the case of two $s$-NLs, the $s$-NLs can be separated, touched or crossed with unrestricted locations in the BZ (see in Fig.(\ref{fig:NL}c)). The crossed $s$-NLs are also called Hopf-link due to their topological invariant being the Hopf-link number.\cite{NLSM_WangHopfPRB2017} While the existence of multiple-loop NLs has been confirmed in photonic lattice,\cite{NodeChain_LuLingNP2018} their existence in fermionic systems has not been identified thus far.

In this work, using first principles calculations, we screened the family of MgSrSi-type crystal which consists of 660 members, and found more than 70 compounds are NLSMs showing a variety of NL structures. The NLs are protected by the mirror symmetry or the $\mathcal{S}$ symmetry contained in the Pnma space group of MgSrSi-type crystals. Importantly, in AsRhTi, we found a new type of multi-loop NL structure as shown in Fig.(\ref{fig:NL}d), where a $s$-NL sticks to a $m$-NL at some ``nexus point" (denoted by $O$) and penetrates the invariant plane of $m$-NL at some general point (denoted by $E$). Interestingly, we also found a third $m$-NL crosses the $s$-NL and a Hopf link is formed by the two. Unlike the case of two intersecting $m$-NLs or two crossed $s$-NLs, this novel multi-loop NL structure requires a minimal three-band model to describe its essential electronic structure. Even more exotic multi-loop NL structures were uncovered in AsFeNb and PNiNb. Some of the NLSMs show very clean band structures at the fermi level without other trivial bands. Our work therefore provides a promising platform for the material realization of new topological semimetals with exotic NL structures.

\section{Methods}
The first-principle calculations were performed by the Vienna \emph{ab initio} simulation package (VASP) \cite{VASP} and projected augmented-wave (PAW) potential is adopted.\cite{PAW_Blochl,PAW_Kresse_1999} The exchange-correlation functional introduced by Perdew, Burke, and Ernzerhof (PBE)~\cite{PBE} within generalized gradient approximation (GGA) is applied in the calculations. The energy cutoff for the plane-wave basis is set as 520 eV and the forces are relaxed less than 0.01 eV/$\rm \text{\AA}$. The positions of atoms are allowed to relax while the lattice constants of the unit cells are fixed to the experimental values documented in the Inorganic Crystal Structure Database (ICSD). The band-crossings are calculated from tight-binding models which are constructed by using the Maximally Localized Wannier Functions (MLWF) method coded in WANNIER90.\cite{Wannier90}

\section{Multi-LOOP NLs in AsRhTi}

\begin{figure*}[htbp]
\centering{}
\includegraphics[width=0.95\textwidth]{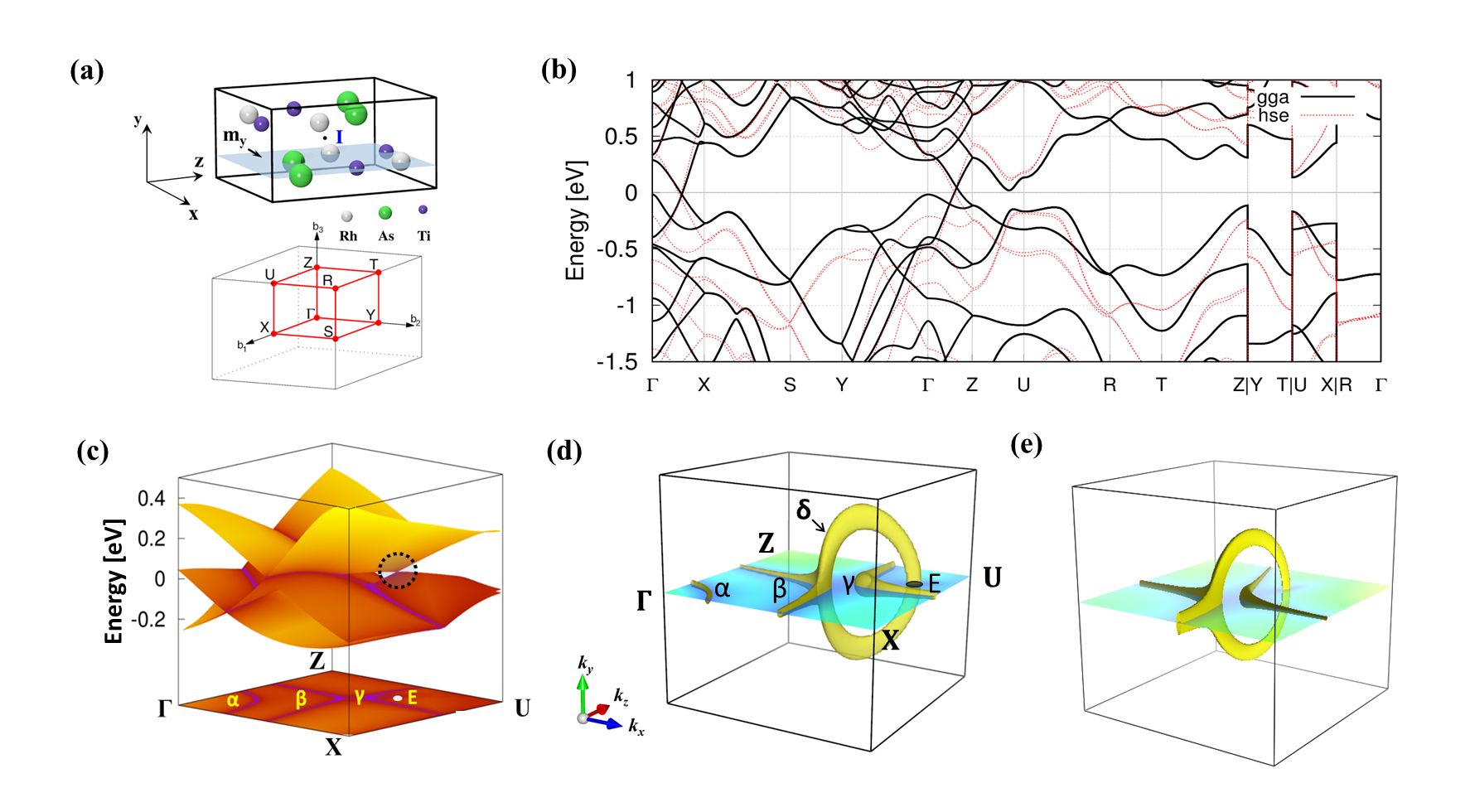}
\caption{(Color online)
NLs band crossings of AsRhTi crystal. (a) The crystal structure and Brillouin Zone of AsRhTi. The crystal takes a space group of Pnma. The mirror plane of $m_y$ and space inversion symmetry $\mathcal{P}$ are highlighted. (b) Band structure of AsRhTi. Both band structures obtained from the GGA and HSE06 calculations are presented by solid and dotted lines, respectively. (c) Band structure on the invariant plane $k_y=0$. Only three bands at the fermi level are plotted and the energy differences between these bands are projected on to the bottom of the figure. Three $m$-NLs $\alpha$, $\beta$ and $\gamma$ are highlighted. The point E denoted an isolated band crossing point on the $k_y=0$ plane. (d) The band crossings in the 3D BZ. The point E is the isolated intersection points of an $i-$NL and the $k_y=0$ plane. (e) The band crossing calculated from the $k\cdot p$ model in Eq.(\ref{eq:Heff2}). The dimensionless parameters are set as $\epsilon_1=-0.346$, $\epsilon_2=-0.865$, $\epsilon_3=-0.26$, $a_1=1.0$, $a_2=1.0$, $a_3=0.15$, $b_1=0.05$, $b_2=1.0$, $b_3=0.15$, $c_1=1.0$, $c_2=0.05$, $c_3=0.15$, $\lambda_1=0.588$ and $\lambda_2=0.588$.
\label{fig:bnd}}
\end{figure*}

The MgSrSi-type crystals take a Pnma space group which contains three mirror symmetries, $m_x$, $m_y$ and $m_z$, and the inversion symmetry $\mathcal{P}$. If the time reversal $\mathcal{T}$ is also a symmetry, the compounds become symmetrical under the composed operation $\mathcal{S=PT}$. Both conditions of existence of the $s$-NL and $m$-NLs are therefore fulfilled in MgSrSi-type crystals. In Fig.(\ref{fig:bnd}a) we plotted the crystal structure of a prototype compound, AsRhTi and the two key symmetries for NLs, a mirror plane $m_y$ and inversion symmetry $\mathcal{P}$ are highlighted. It can be found that in the unit cell of AsRhTi it contains two layers of atoms and the plane of the atom layers is overlapped with the mirror plane of $m_y$. The inversion center, on the other hand, is off the atom layers and locates at the middle of two neighboring layers. The corresponding Brillouin Zone and high-symmetry paths are also shown below the crystal structure of AsRhTi.

Let us show here that the valence and conduction bands of AsRhTi do cross and produce a multi-loop NL structure. We plotted the GGA band structure of AsRhTi in Fig.(\ref{fig:bnd}b) where one can readily find that the valence and conduction bands cross at the intermediate points of $\Gamma$-X and $\Gamma$-Z, indicating a NL lying in the invariant plane $k_y=0$ of the mirror symmetry $m_y$. In order to demonstrate NLs more clear, we further plotted in Fig.(\ref{fig:bnd}c) the 2D band structure of AsRhTi on the $k_y=0$ plane with varying $k_x$ and $k_z$. Since the band structure is symmetrical under the transformations $k_x\rightarrow -k_x$ and $k_z\rightarrow -k_z$, only the region of $k_x>0$ and $k_z>0$ is used for simplicity. From the distribution of energy difference of bands that has been projected on the bottom of Fig.(\ref{fig:bnd}c), one sees a central NL $\alpha$ surrounding the $\Gamma$ point. Outside the NL $\alpha$ is a $m$-NL $\beta$. Outmost is a third $m$-NL $\gamma$ that encloses the U point. It is the $m_y$ provides the needed protection for these $m$-NLs. Interestingly, besides these in-plane $m$-NLs, we also find an isolate band-touching point E outside the NL $\gamma$. This band-touching point E is more obviously seen in the 2D energy band structure where it is highlighted by a dotted circle in Fig.(\ref{fig:bnd}c).

The detail of the NLs near the point $E$ is revealed by a 3D profile of energy differences of bands in Fig.(\ref{fig:bnd}d), where a denser discretion of BZ is adopted to obtain the energy bands with the tight-binding hamiltonian constructed by the MLWF method. In this band-crossing profile, an extra NL $\delta$ vertical to $k_y=0$ plane is discovered. Since NL $\delta$ does not lie in any high symmetry path or plane, it must be an $s$-NL that is protected by symmetry $\mathcal{S}$. More detailly, the $s$-NL $\delta$ is found to stick to the $m$-NL $\beta$ on a nexus point and the point E is the very point that NL $\delta$ penetrates through the $k_y=0$ plane. Another interesting feature in Fig.(\ref{fig:bnd}d) is that the $m$-NL $\gamma$ crosses the $s$-NL and the two form a Hopf-link. Previously Hopf-link made of two $m$-NLs or two $s$-NLs have been already proposed and it is argued that for the first one needs a minimal four-band effective model to describe the electronic structure of the Hopf-link,\cite{NLSM_CoMnGaPRL2017}, while for the later one only needs a minimal two-band model.\cite{NLSM_WangHopfPRB2017,NLSM_ChenHopfPRB2017} Here we revealed that a Hopf-link can be made of a $s$-NL and a $m$-NL, and its corresponding electronic structure is correctly described by a minimal three-band model given below. It is indeed seen in Fig. (\ref{fig:bnd}b) that the band structure on the $\Gamma$-X and $\Gamma$-Z involves three bands around the Fermi level.

The reason that we need a minimal three-band model to describe the multi-loop NL structure of Fig.(\ref{fig:bnd}d) is obvious: The $m$-NL is only produced by a pair of bands with opposite mirror parities. A robust and isolate band crossing point E on the invariant plane $k_y=0$, however, is only possible when the crossing bands have equal mirror parities. Otherwise there would be a $m$-NL passing through the isolate point. \cite{NLSM_YuHfCPRL2017}
The general form of the three-band hamiltonian should be written as,
\begin{eqnarray}
H(\bf{k})=\left[ \begin{array}{ccc}
                         H_{11}(\bf{k}) & H_{12}(\bf{k}) & H_{13}(\bf{k})  \\
                           & H_{22}(\bf{k}) & H_{23}(\bf{k}) \\
                         \dag &  & H_{33}(\bf{k})
                       \end{array}
\right].
\label{eq:Heff1}
\end{eqnarray}
Since the system preserves the symmetry $\mathcal{S}$, the imaginary part of the off-diagonal element $H_{nm}(\bf{k})$ ($n,m=1,2,3$ and $n\neq m$) vanishes. On the other hand, The mirror symmetries, i.e. $m_x$, $m_y$ and $m_z$, lay on the entries another constraint that the diagonal element $H_{nn}(\bf{k})$ should be an even function of $k_x$, $k_y$ and $k_z$.\cite{NLSM_YuPRL2015} For the off-diagonal entry $H_{nm}(\bf{k})$ with $n\neq m$, it becomes an even (odd) function of $k_i$ ($i=x,y,z$) if the orbital $n$ and $m$ have the equal (opposite) mirror parities with respect to symmetry $m_i$.\cite{NLSM_YuPRL2015} The above symmetry consideration helps us to reduce the hamiltonian of Eq. (\ref{eq:Heff1}) to a simpler form up to a second order of $k$,
\begin{widetext}
\begin{eqnarray}
H(\bf{k})=\left[ \begin{array}{ccc}
                         0 & \lambda_1 k_y & \lambda_2 k_y  \\
                           & \epsilon_{1}+a_1 k_x^2+b_1k_y^2+c_1k_z^2 & \epsilon_{3}+a_3 k_x^2+b_3k_y^2+c_3k_z^2 \\
                         \dag &  & \epsilon_{2}+a_2 k_x^2+b_2k_y^2+c_2k_z^2
                       \end{array}
\right].
\label{eq:Heff2}
\end{eqnarray}
\end{widetext}
 Here we supposed that the second and third orbitals have the equal mirror parities opposite to the first orbital. A constant term $H_{11}\hat{I}$ has been subtracted from the original Hamiltonian because of its irrelevance to the structure of NLs. The parameters of Eq.(\ref{eq:Heff2}) are chosen dimensionless for simplicity. By choosing suitable values of parameters $\epsilon_i$, $a_i$, $b_i$, $c_i$ and $\lambda_j$ ($i=1,2,3$ and $j=1,2$), the main features of multi-loop NL structure of Fig.(\ref{fig:bnd}d) are well reproduced as shown in Fig.(\ref{fig:bnd}e).

\begin{figure}[htbp]
\centering{}
\includegraphics[width=0.48\textwidth]{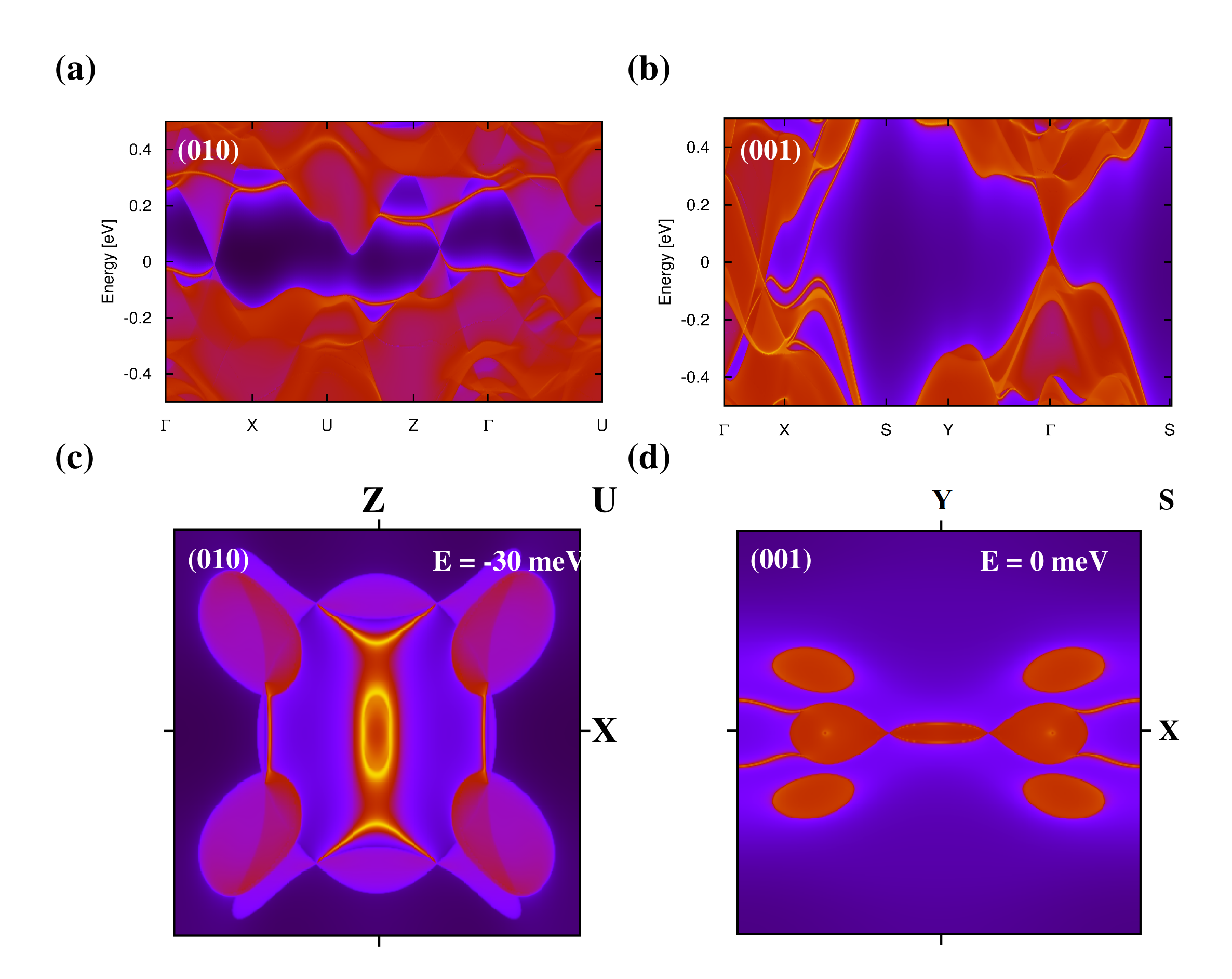}
\caption{(Color online)
Surface electronic structure of AsRhTi. (a) and (b) show the surface band structures on the (010) and (001) surfaces, respectively. (c) and (d) plot the 2D surface DOS profiles corresponding to (a) and (b). The energies of (c) and (d) are set as -30 and 0 meV, respectively.
\label{fig:ss}}
\end{figure}

The nontrivial electronic structure of a NLSM is revealed by its topological surface states (SSs). In a slab, the projection of a NL in the bulk BZ onto the 2D BZ of the slab will divide it into regions of different topological orders characterized by $\mathcal{Z}_2$ topological charge $\nu$,
\begin{eqnarray}
\nu=\frac{1}{\pi}\int_{-\pi}^{\pi} dk_{\perp} \sum_{n\in occ.}\langle n,\textbf{k}| \emph{i} \partial_{\emph{k}_{\perp}}|n,\textbf{k}\rangle \mod 2,
\label{eq:nu}
\end{eqnarray}
where $|n,\textbf{k}\rangle$ denotes the Bloch eigenstate and $k_{\perp}$ is the component of momentum normal to the slab. In the regions of $\nu=1$, there exists in-gap topological surface states at each k point, forming the so called 2D ``drum-head" states.\cite{YXZhao_arxiv2017}
In Fig.(\ref{fig:ss}a) and (\ref{fig:ss}b) we have shown the surface band structures of AsRhTi on the (010) and (001) surfaces, respectively. The corresponding 2D profiles of density of state (DOS) at fixed energies of -30 and 0 meV are also plotted in Fig.(\ref{fig:ss}c) and (\ref{fig:ss}d). In Fig.(\ref{fig:ss}a) and Fig.(\ref{fig:ss}c), one finds SSs spreading throughout the inner region enclosed by the projection of $m$-NL $\beta$ of Fig.(\ref{fig:bnd}d). Since the $s$-NLs are normal to surface (010) (see in Fig.(\ref{fig:bnd}d)), no SS is found at (010) surface for the $s$-NLs. In contrast, on the (001) surface the projection of $s$-NLs form two ellipses on $\Gamma$-X and SSs link the two ellipses across the boundary of BZ (see in Fig.(\ref{fig:ss}d)). It is seen from Fig.(\ref{fig:ss}d) that the fermi surface also cuts some trivial bands, producing carrier pockets above and below the $k_y=0$ plane.

\section{Diverse Node lines structure in MgSrSi-type crystals}

\begin{figure*}
\centering{}
\includegraphics[width=0.9\textwidth]{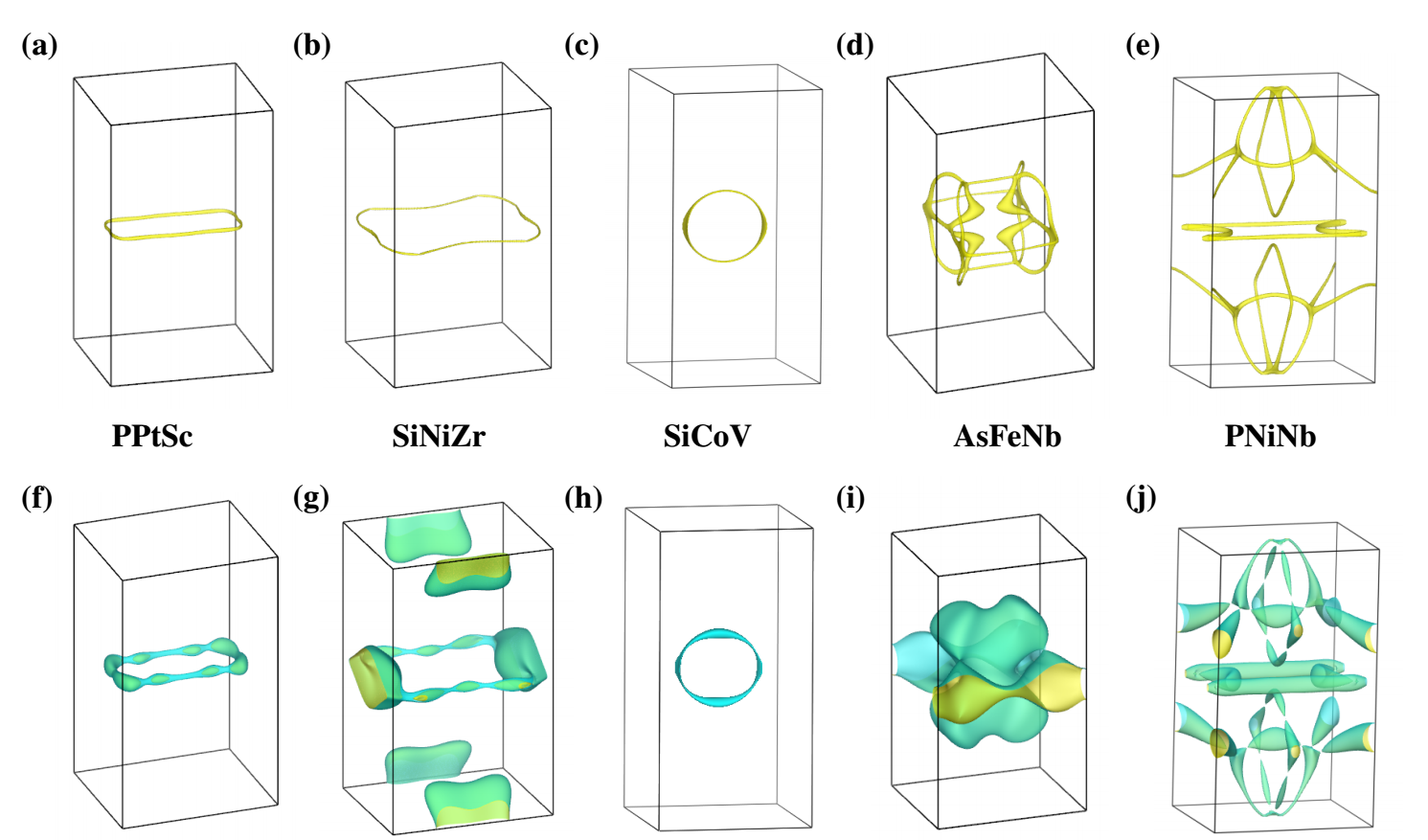}
\caption{(Color online)
Diverse NLs in MgSrSi-type ternary crystals. (a)-(e) plot the band crossings of AsFeNb, PPtSc, SiNiZr, SiCoV and PNiNb. (f)-(j) show the fermi surfaces of AsFeNb, PPtSc, SiNiZr, SiCoV and PNiNb.
\label{fig:NL2}}
\end{figure*}
\begin{table}[htbp]
\caption{NLSMs in MgSrSi-type crystals}
\renewcommand{\arraystretch}{1.4}
	\begin{tabular}{p{0.25\columnwidth}|p{0.7\columnwidth}}
	\hline
	\hline
	Catalog & Node-Line Semimetals \\
	\hline
	AsRhTi$^G$ & AsRhTi, AsCoHf, PCoHf, PCoTi, PCoZr, PRhZr \\
	\hline
    PFeV$^G$  & PFeV, PFeNb, PFeTa, AsFeNb, AsFeTa, PRuNb, PRuTa\\
	\hline
	PNiV$^G$  & PNiV, PNiNb, PNiTa, AsNiV, AsNiNb, AsNiTa\\
	\hline
	PPtSc$^G$ & PNiSc, PPtSc, PPdEr$^*$, AsPdLn$^*$(Ln=La, Gd, Ho, Tb, Dy, Tm) \\
	\hline
	SiCoV$^G$ & SiCoV, SiCoNb, SiCoTa, SiRhNb, SiRhTa, SiIrNb, SiIrTa, GeCoV, GeCoNb, GeCoTa, GeRhNb, GeRhTa, GeIrTa\\
	\hline
	SiNiTi$^G$ & SiNiTi, SiNiZr, SiNiHf, SiPdTi, SiPdZr, SiPdHf, SiPtTi, SiPtZr, SiPtHf, GeNiZr, GeNiHf, GePdTi, GePdZr, GePtZr, GePtHf\\
	\hline
     other$^G$ & GeBa$_2$, SiBa$_2$, PbBa$_2$, PbCa$_2$, SnBa$_2$, SiMgBa, SiMgCa, SiMgSr, SiCaBa, GeSrMg, SnCaMg, SrMgSn, CaBiLi\\
	\hline
	\hline
\end{tabular} \\
$^*$ f-electrons are not included in the calculations.
\label{tab:NLSM}
\end{table}

The MgSrSi-type crystals is a large family of ternary crystals which contains more than 660 members. As expected that iso-structure crystals may have similar electronic structure, we thus screened all 660 compounds' to discover new NLSMs. More than 70 NLSMs are readily found and listed in Tab.(\ref{tab:NLSM}), where the NLSMs are divided into several groups based on their chemical compositions. A variety of NLs structures were discovered. In Fig.(\ref{fig:NL2}) we plotted 5 representative NL structures and their corresponding fermi surfaces. From Fig.(\ref{fig:NL2}a), one sees PPtSc has a single NL loop lying in the $k_y=0$ plane and its fermi surface takes a distorted torus-like shape (see in Fig.(\ref{fig:NL2}f)). For SiNiZr its NL extends across the boundary of BZ and one can see from Fig.(\ref{fig:NL2}b) that some portions of the NL outside the BZ is fold back. The single NL of SiCoV shown in Fig.(\ref{fig:NL2}c), contrarily, lies on the $k_z=0$ plane unlike those of PPtSc and SiNiZr in the $k_y=0$ plane, indicating its protecting symmetry being $m_z$. Both fermi surfaces of PPtSc and SiCoV take the simple torus-like shapes and show very clean fermi surfaces, a promising property for the experimental detection of their nontrivial electronic structures.

Interestingly, even more exotic NL geometries are found in AsFeNb and PNiNb. There exist multiple $m$-NLs in the $k_x=0$, $k_y=0$ and $k_z=0$ planes, together with $s$-NLs sticking to the $m$-NLs (see in Fig.(\ref{fig:NL2}d) and Fig.(\ref{fig:NL2}e)). For AsFeNb, its NLs form a novel cage-like structure. However, its fermi surface is dirty which is messed up by some trivial bands. For PNiNb, one finds an isolated $m$-NL lies in the invariant plane $k_y=0$, and off the plane NLs protected by $m_x$ and $m_z$ mirror symmetries are found touching at Y point. Around the planes of $k_x=\pm k_z$, there exist eight segments of $s$-NLs sticking to he $m$-NLs of $k_x=0$ plane. Luckily the fermi surface of PNiNb, shown in Fig.(\ref{fig:NL2}j), is very clean and quite similar to the NLs structure of Fig.(\ref{fig:NL2}e). Therefore PNiNb can be a very interesting compound for exploring new NLSMs with exotic multi-loop NL structure.

\section{Discussions}
In this work, the study are based on the nonmagnetic ground states of the compounds. At room temperature, this assumption may be reasonable for magnetic compounds with low Curie or Neel temperatures. It is worthy to note that for many MgSrSi ternary compounds which contain magnetic transition metal elements the ground states are actually non-magnetic. For example, the ternary MM'X (M=transition metal, M'= late transition metal, X= main group element) compounds usually exhibit paramagnetic behaviors for M = Sc, Ti and V.\cite{MMX-MAG} Previously, TiCoP, ZrCoP and VCoSi have already been characterized to be paramagnetic metallic conductors.

In all our calculations, we have not implemented the SOC in our calculations. For many crystals such as AsRhTi and PNiNb  etc., the SOC induces small gaps on the NLs at the scale of several meV to ten meV. Therefore at room temperature the effective of SOC is ignorable. However, for compounds containing heavier elements, such as SiIrTa, the SOC gap is not small that the NLSMs eventually are turned into topological insulators. we also check effect of electron interactions by adopting the hybrid density functional approximation (HSE06)\cite{HSE06} and find that most of the NLSMs are found to retain their inverted valence and conduction bands even under the hybrid density functional approximation. For example, we plotted the HSE06 energy bands in the Fig.(\ref{fig:bnd}b), from which one finds that the valence and conduction bands are kept inverted.

We note some of the NLSM compounds have already been reported previously.\cite{NLSM_WTLHfPtGe2018,AsRhTiPRB2018,ZrPtGePRB2018} Especially, AsRhTi has been proposed to be a NLSM.\cite{AsRhTiPRB2018} However, only the single NL in the plane $k_y=0$ is revealed by the authors but not the novel multi-loop structure in Fig.(\ref{fig:bnd}d). The reason may be that since $i$-NL is unpinned to any high-symmetry path or plane, it could be missed in a plotting of band structures along high symmetry paths. Only calculations performed on denser discretion of bulk BZ can uncovered such $i$-NL off the high symmetry path or plane.

\vspace{3mm}
\noindent\textit{Acknowledgements---}
The work is supported by National Natural Foundation of China (NFSC) (Grants No.11574215, No. 11274359 and No. 11422428).  H. M. W is also supported by the National 973 program of China (Grants No. 2011CBA00108 and No. 2013CB921700), and the ``Strategic Priority Research Program (B)" of the Chinese Academy of Sciences (Grant No. XDB07020100). The calculations in this work were performed on the supercomputers of Shanghai supercomputer Center and of the high performance computing center of Nanjing University.
J. L. and L. Y. contributed equally to this work.

\bibliography{refs}
\end{document}